\documentclass[a4paper,11pt]{article}

\usepackage{jinstpub} 
\usepackage[squaren]{SIunits}
\usepackage{lineno}

\usepackage[normalem]{ulem}
\usepackage{url} 
\usepackage{caption}
\usepackage{subfig, graphicx} 
\usepackage{floatrow}
\makeatletter

\newcommand{\Rmnum}[1]{\expandafter\@slowromancap\romannumeral #1@}
\makeatother

\title{\boldmath MUX64, an analogue 64-to-1 multiplexer ASIC for the ATLAS High Granularity Timing Detector}

\author[a,1]{Z.~ XU,\note{\textcolor{black}{Zifeng XU and Li ZHANG are considered co-first authors.}}}
\author[b,1]{L.~ ZHANG,}
\author[b]{X.~ HUANG,}
\author[c,d]{Q. SHA,}
\author[a]{Z.~ GE,}
\author[a]{Y.~ CHE,}
\author[b]{D.~ GONG,}
\author[e]{S.~ HOU,}
\author[c,d]{J.~ ZHANG,}
\author[b]{T.~ LIU,}
\author[c,d]{Z.~ LIANG,}
\author[a]{L.~ ZHANG,}
\author[b]{J.~ YE,}
\author[a,2]{M.~ QI\note{\textcolor{black}{Corresponding author.}}}

\affiliation[a]{\textcolor{black}{Department of Physics, Nanjing University, Nanjing 210093, China}}
\affiliation[b]{\textcolor{black}{Department of Physics, Southern Methodist University, Dallas TX 75275, USA}}
\affiliation[c]{\textcolor{black}{Institute of High Energy Physics, Chinese Academy of Sciences, Beijing 100049, China}}
\affiliation[d]{\textcolor{black}{University of Chinese Academy of Sciences, Beijing 100040, China}}
\affiliation[e]{\textcolor{black}{Institute of Physics, Academia Sinica, Taipei 11529, Taiwan}}

\emailAdd{qming@nju.edu.cn}

\abstract{We present the design and the performance of MUX64, a 64-to-1 analogue multiplexer ASIC for the ATLAS High Granularity Timing Detector (HGTD). The MUX64 transmits one of its 64 inputs selected by \textcolor{black}{six address lines} for the voltages or temperatures being \textcolor{black}{monitored} to an lpGBT ADC channel. The prototype ASICs fabricated in TSMC $ 130\,\nano\meter $ CMOS technology were prepared in wire-bonding and QFN88 \textcolor{black}{packaging format}. \textcolor{black}{A} total of 280 chips \textcolor{black}{was} examined for functionality and quality assurance. The accelerated aging test conducted at $ 85\degreecelsius $ shows negligible degradation over \textcolor{black}{16 days}.}


\proceeding{topical workshop on electronics for particle physics 2022\\
	19-23 September, 2022\\
	Bergen, Norway}

\begin{document}
	\maketitle
	\flushbottom
	\section{Introduction}
	\label{sec:introduction}
	The High-Luminosity Large Hadron Collider (HL-LHC) at CERN aims for delivering an integrated luminosity up to 4000 $ \femto\barn^{-1} $. The instantaneous luminosity will be increased to $ 7.5 \times 10^{34}\,\centi\meter^{-2}\second^{-1} $, \textcolor{black}{which is} a factor 3-4 to the Run2 of the LHC. The event pileup caused by higher luminosity is one of the main challenges at the HL-LHC. The High Granularity Timing Detector (HGTD) \cite{CERN-LHCC-2020-007} \textcolor{black}{of the ATLAS Phase-\Rmnum{2} upgrade \cite{ATLAS_1502664}} is in construction to mitigate the pileup effects. 
	
	\textcolor{black}{The HGTD detector module consists of a Low-Gain Avalanche Detectors (LGAD) \cite{PELLEGRINI201412} sensor bump-bonded to two readout chips (ALTIROC \cite{9507972}) for a timing resolution of $ 25\,\pico\second $.} In the peripheral area surrounding the detector modules, the Peripheral Electronics Boards (PEBs) transmit the data between the front-end detector modules and the data acquisition (DAQ) system, 
	the detector control system (DCS) \textcolor{black}{and the luminosity systems}. The \textcolor{black}{detector} modules are connected to the PEBs via flexible circuit cables \textcolor{black}{(FLEX)} \cite{Robles_Manzano_2022}. It is important to monitor the \textcolor{black}{temperatures} of the LGAD sensors and the supply voltage drops in FLEX cables. 
	 \textcolor{black}{The analogue signals being monitored are read by the ADC of the Low Power Giga Bit Transceiver (lpGBT \cite{GBT_project}) mounted on the PEBs.}
	
	To accommodate \textcolor{black}{the} large number of \textcolor{black}{monitored} channels for the detector modules, the design incorporating a multiplexer is necessary for readout by a single ADC channel on an lpGBT. The MUX64, a 64-to-1 analogue multiplexer has been developed to meet the requirements. \textcolor{black}{Six GPIO output ports of the lpGBT are connected to the MUX64 to select an input signal transferred to the output.} \textcolor{black}{Compared} to commercial multiplexers, the MUX64 handles more (up to 64) inputs. 
	 \textcolor{black}{The total of 1300 MUX64 chips will be produced for the HGTD, which include 15\% spare.}

	This paper is organized as the following: in section \ref{sec:mornitor_mod} we describe the monitoring system for the detector modules of the HGTD. The design of MUX64 and the requirements are discussed in Section \ref{sec:chip_design}. The prototype MUX64 chips were \textcolor{black}{examined} and the results are presented in section \ref{sec:measurements}. A summary and outlook on radiation test are discussed in \ref{sec:conclusions}.

	\section{Monitoring of the \textcolor{black}{HGTD} detector modules}
	\label{sec:mornitor_mod}

%
	\begin{figure}[htb]
	\centering
	\includegraphics[width=0.6\linewidth]{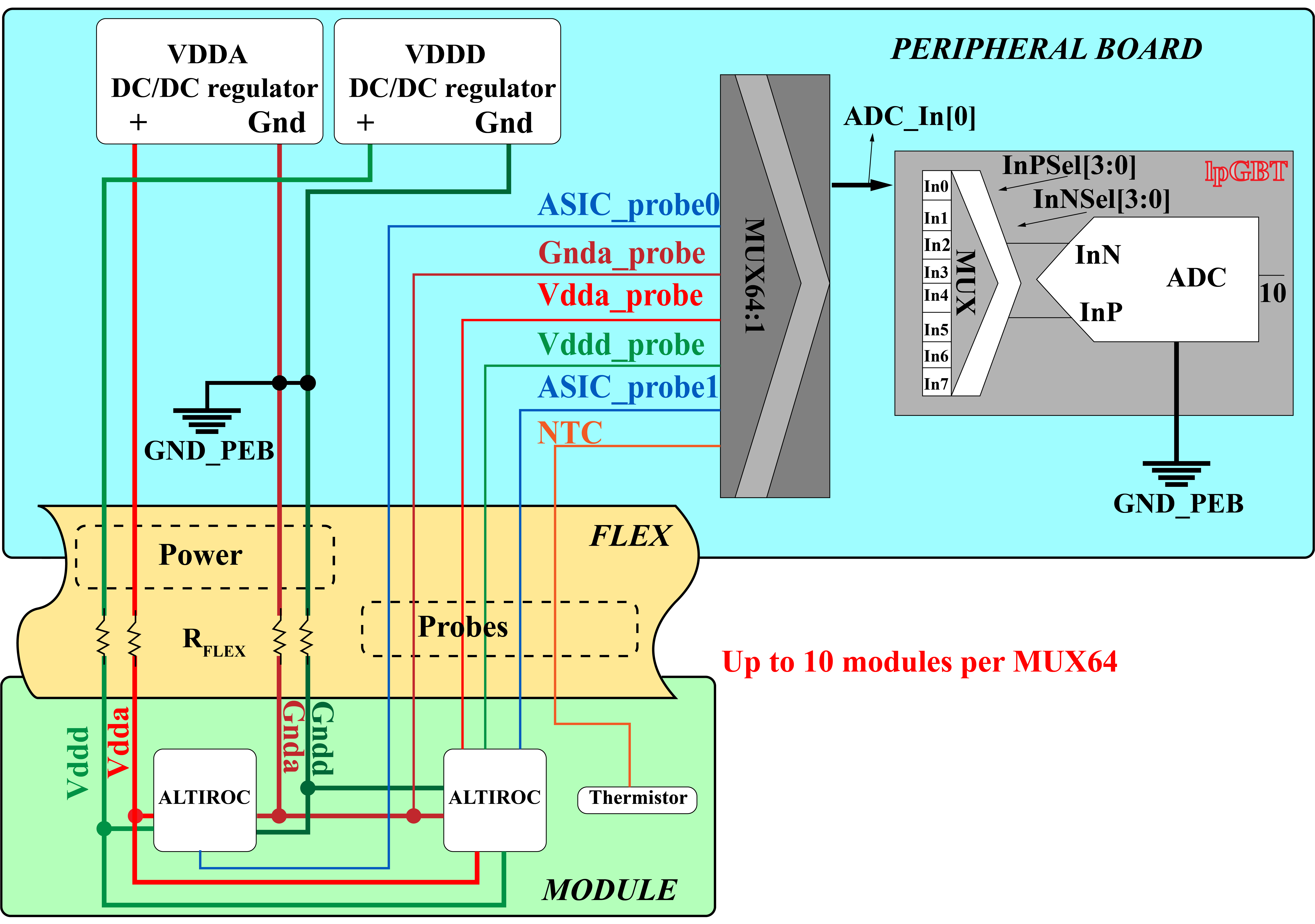}
	\caption{Schematic \textcolor{black}{view} of the monitoring system: signals are collected by a MUX64 multiplexer to an ADC on the lpGBT \cite{CERN-LHCC-2020-007}.}
	\label{fig:monitoring_schematic}
	\end{figure}
	The digital and analogue supply voltages on the \textcolor{black}{HGTD} detector \textcolor{black}{modules} and \textcolor{black}{the} DC/DC regulators shall be monitored for operating stability and the supply voltage drops \textcolor{black}{passing through} the FLEX cable. Detecting voltage change at the module level is an effective method for finding \textcolor{black}{possible latch-up events} on an ALTIROC. 
	\textcolor{black}{The LGAD sensors which are sensitive to temperature will be operated at $ -30\,\degreecelsius $ to reduce leakage current.} The temperature sensors 
	are implemented inside the ALTIROC chips which provide a resolution of $ 0.2\,\degreecelsius $ to detect \textcolor{black}{loss of cooling} and thermal runaway in a range from $ -40 $ to $ +40\,\degreecelsius $.

	 Figure \ref{fig:monitoring_schematic} shows the schematic of the monitoring of a detector module. A total of 6 signals are \textcolor{black}{sent to} the MUX64. Three of them are \textcolor{black}{the} supply voltages and analogue grounding ($ \volt_{\mathrm{ddd\_probe}} $, $ \volt_{\mathrm{dda\_probe}} $ and $ \volt_{\mathrm{Gnda\_probe}} $), another two are \textcolor{black}{the} analogue signals from 2 ALTIROC ($\mathrm{ASIC_{probe0}}$, $\mathrm{ASIC_{probe1}}$), and one from thermistor (NTC). \textcolor{black}{These signals are stable low-frequency analogue signals. The expected switching frequency \textcolor{black}{in the monitoring system} is about $ 100\,\hertz $. This is below the MUX64 specification of $ 1\,\kilo\hertz $.} \textcolor{black}{Up to 10 detector modules can be monitored by a MUX64 to an ADC channel of a single lpGBT.}

	\section{MUX64 Chip design}
	\label{sec:chip_design}
	\subsection{Specifications}
	\label{subsec:design_spec}
	The MUX64 is \textcolor{black}{designed} for processing 64 analogue inputs to one output. The operating temperature range is $ -35\,\degreecelsius $ to $ +40\,\degreecelsius $. The power consumption of a MUX64 should be smaller than $ 1\,\milli\watt $. The input dynamic range is $ 0.0 $ to $ 1.0\volt $, and the output is read by the 10-bit ADC of the lpGBT.
	
	The input impedance of the ADC is evaluated to be \textcolor{black}{about $ 4.8\,\mega\ohm $.} $ R_{\mathrm{ON}} $ of the MUX64 is the resistance between selected "ON" input channels and the output, which is recommended to be lower than $ 900\,\ohm $ for matching the precision of the ADC. The $ R_{\mathrm{OFF}} $ between the "OFF" channels and the output is required to be larger than $ 60\,\mega\ohm $. Due to the limited space on PEBs, the MUX64s are packaged in the miniature QFN88 format with a size of \textcolor{black}{$ 1\,\centi\meter\times1\,\centi\meter $}.
	
	The MUX64 on the HGTD PEBs will be located at a radial distance near $ 700\,\milli\meter $ from the beam pipe \textcolor{black}{center}. Radiation tolerance is one of the most important requirements. For the HL-LHC operation period, the MUX64 shall be able to \textcolor{black}{withstand} a total ionizing dose (TID) of $ 0.5\,\mega\gray $ and the $ 1\,\mega\electronvolt $ equivalent neutron fluence of $ 1.5\times10^{15}\,(\mathrm{Si})\, n_{\mathrm{eq}}/\centi\meter^{2} $. 
	
	\subsection{MUX64 schematic and logic}
	\label{subsec:design_sche}
	\begin{figure}[htb]
		\centering
		\subfloat[]{\includegraphics[height=0.35\linewidth]{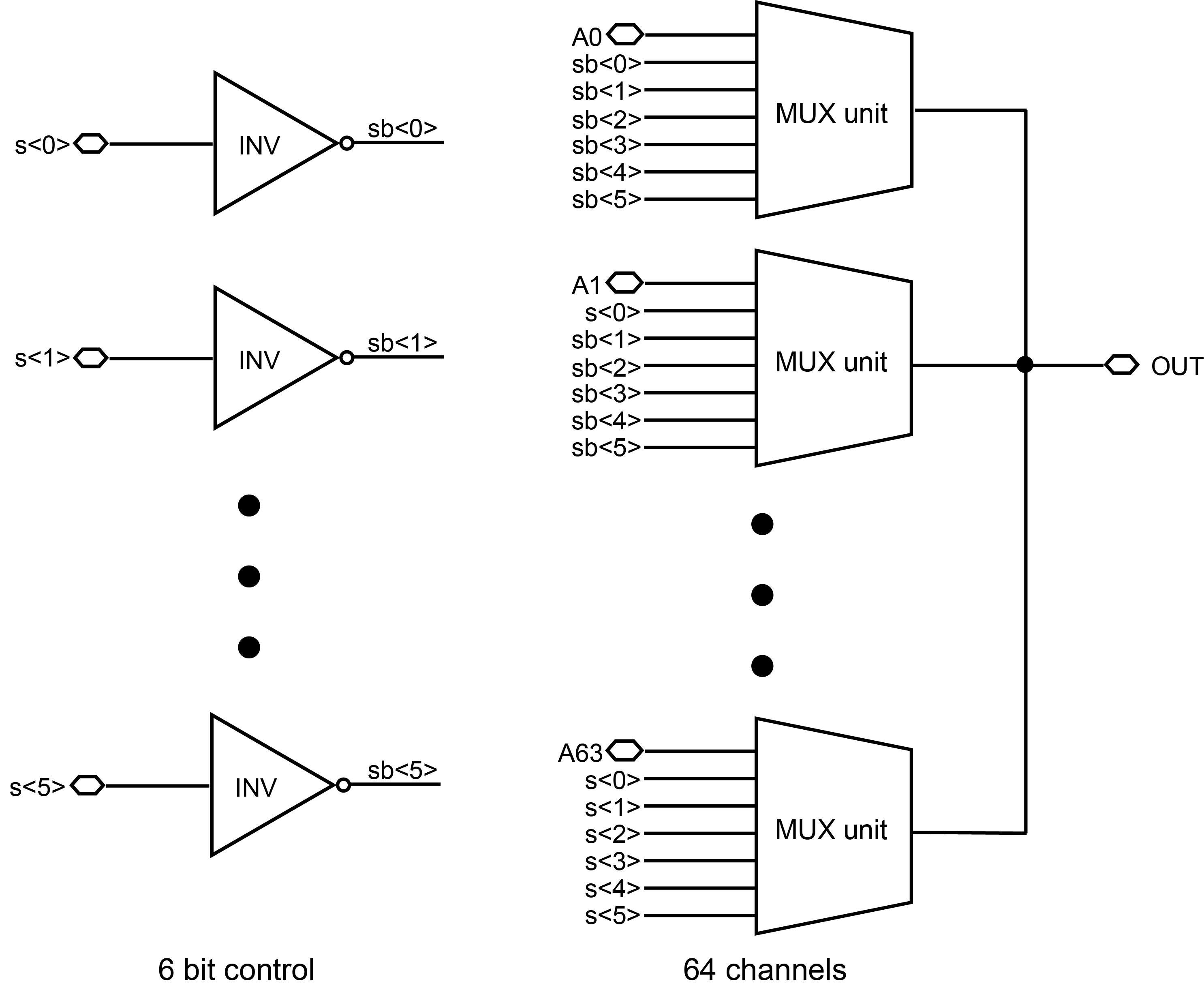}\label{fig:posteraidraft}}\quad
		\subfloat[]{\includegraphics[height=0.35\linewidth]{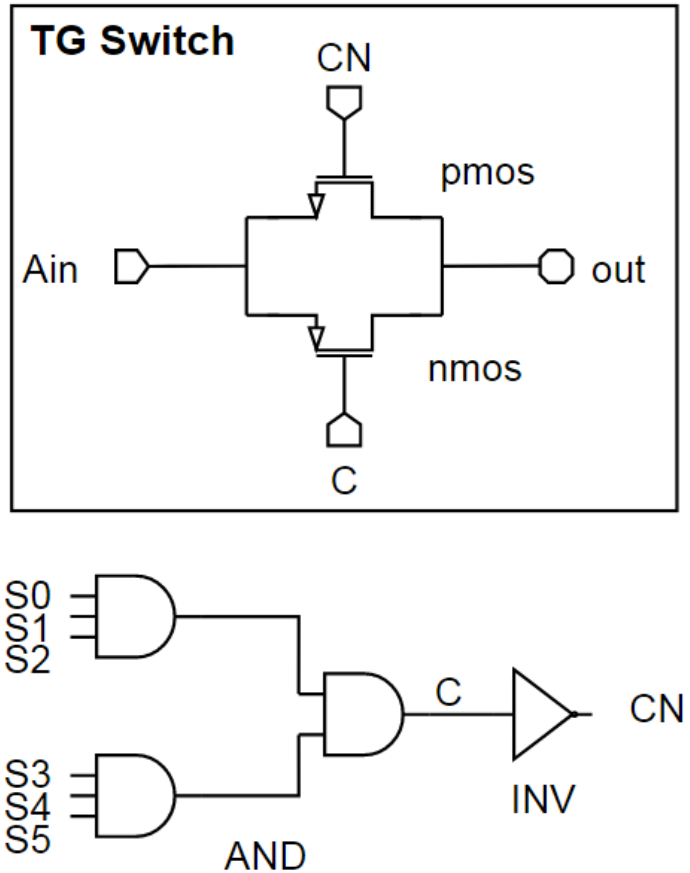}\label{fig:mux64schameticunit}}\\
		\caption{The schematics of (a) the MUX64 design block diagram; and (b) the MUX64 unit with the transmission gate controlled by C and CN.}
		\label{fig:mux_design_schematic}
	\end{figure}
	The block diagram of the MUX64 is shown in Figure \ref{fig:posteraidraft}. The MUX64 uses transmission gates as analogue \textcolor{black}{switches} to transmit only one of the 64 input signals to the output. \textcolor{black}{In the transmission gate circuit, the dimension of PMOS is $ \mathrm{L}=130\,\nano\metre,\,\mathrm{W}=10\,\micro\metre$ and the dimension of NMOS is $ \mathrm{L}=130\,\nano\metre,\,\mathrm{W}=8\,\micro\metre $.} A 6-bit decoder is implemented inside the MUX64 to determine the channel to be connected to the output. The decoder controls the switches of the transmission gates.
	
	The MUX64 is designed and manufactured using the TSMC 130nm CMOS technology. The layout with Enclosed Layout Transistors (ELTs) is employed to enhance radiation tolerance. The decoder is implemented with Triple Modular Redundancy (TMR) structure for fault tolerance against Single Event Upset caused by radiation.

	\section{MUX64 Chip Test}
	\label{sec:measurements}
	\begin{figure}[htb]
		\centering
		\subfloat[]{\label{fig:wire_bonding}\includegraphics[width=0.18\linewidth]{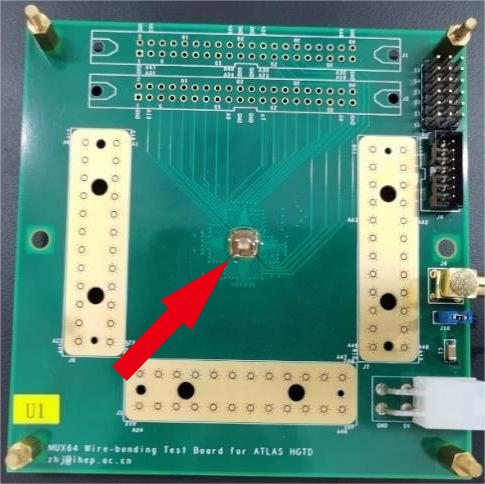}}\quad
		\subfloat[]{\label{fig:QFN_test_socket}\includegraphics[width=0.18\linewidth]{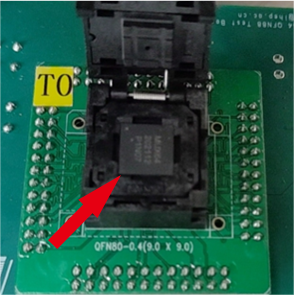}}\quad
		\subfloat[]{\label{fig:em_relay}\includegraphics[width=0.18\linewidth]{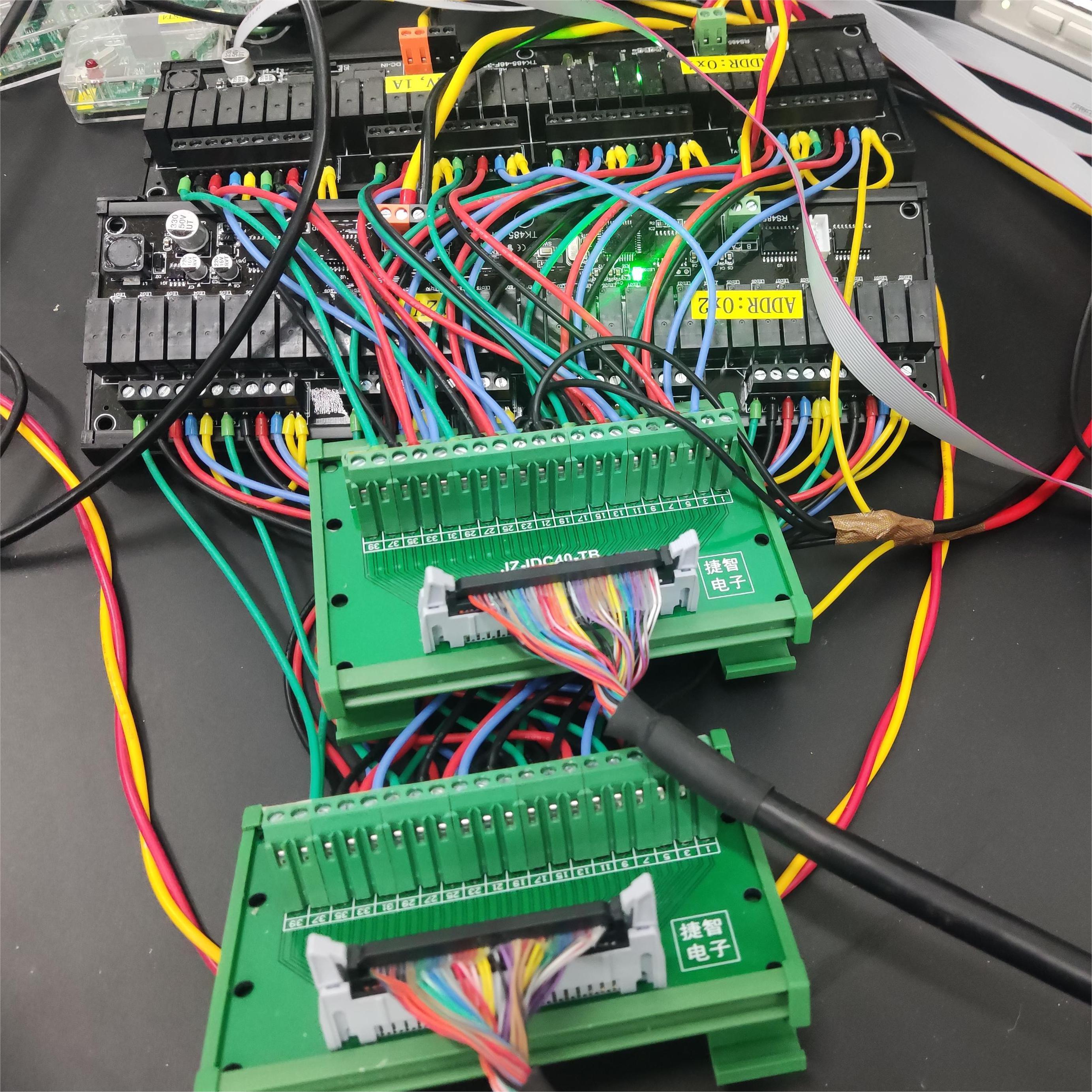}}\quad
		\subfloat[]{\label{fig:photo_UPL}\includegraphics[width=0.18\linewidth]{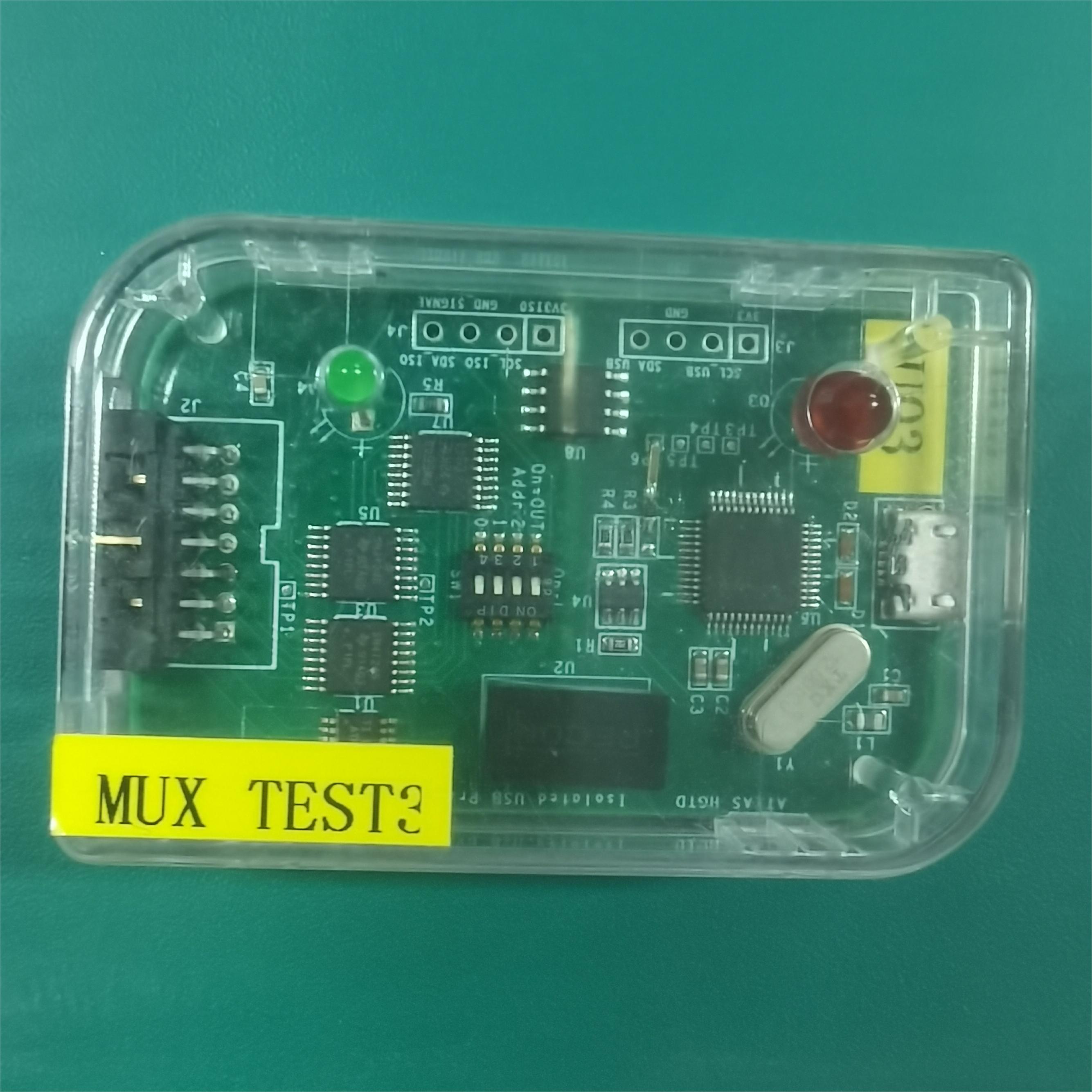}}\quad
		\caption{The MUX64 \textcolor{black}{test kits} are shown for (a) a wire-bonded bare die, (b) a test socket for the QFN88 chips, (c) \textcolor{black}{the 64} electromagnetic relays, (d) the UPL used to select ON-channel for MUX64.}
		\label{fig:MUX64_testing_photos}
	\end{figure}
	 A total of \textcolor{black}{$ 276 $} dies were prepared for characteristic studies.  Twelve dies were wire-bonded on test boards for performance in \textcolor{black}{temperature} \textcolor{black}{change} ranging from $ -40\,\degreecelsius $ to $ 80\,\degreecelsius $. The others \textcolor{black}{were} packaged in QFN88 format \textcolor{black}{which} were examined for quality assurance. The MUX64 test kits are shown in figure \ref{fig:MUX64_testing_photos}. The packaged chips were tested channel by channel with the input voltage \textcolor{black}{ramping} from $ 0.0 $ to $ 1.2\volt $.
	\begin{figure}[htb]
		\centering
		\includegraphics[width=0.8\linewidth]{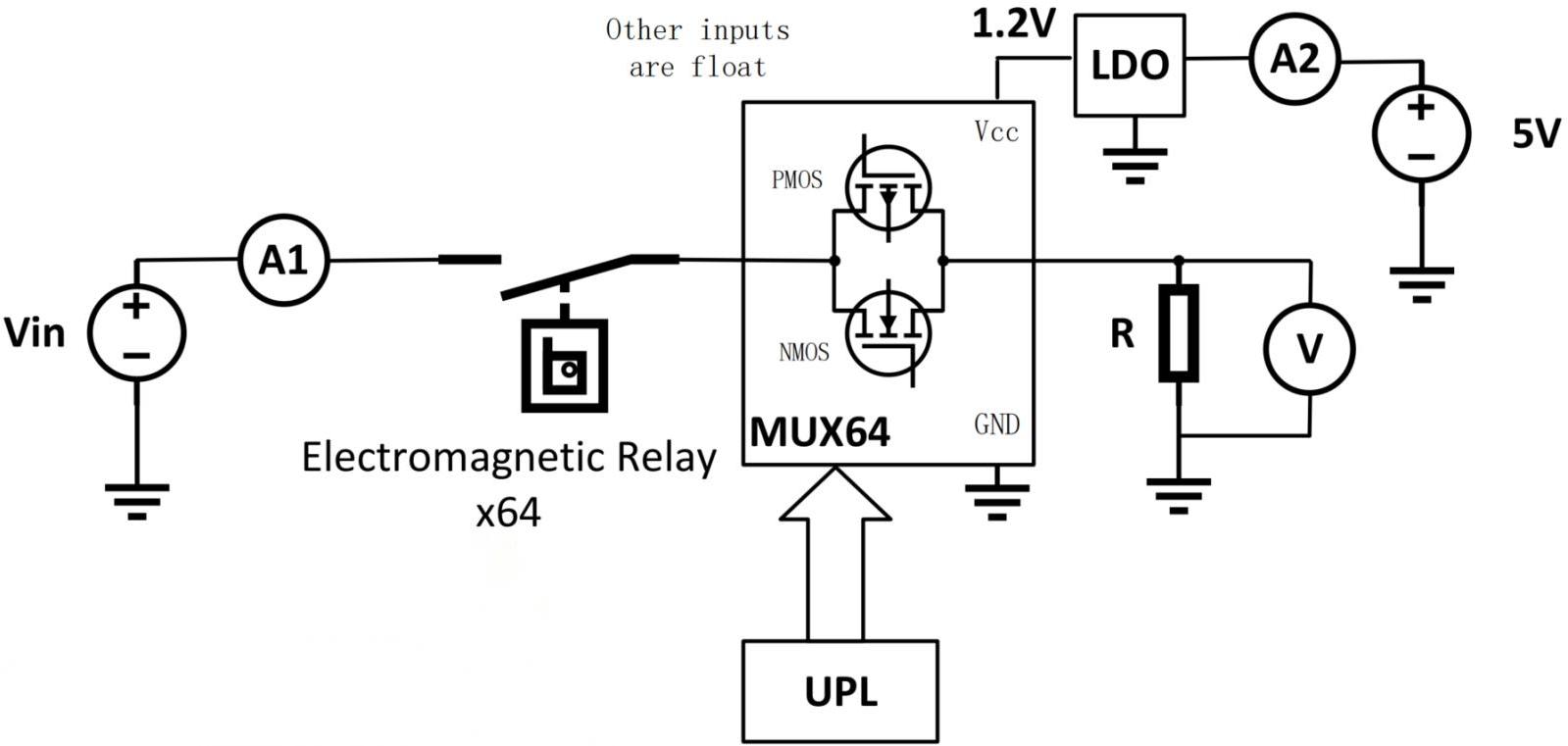}
		\caption{Schematic of \textcolor{black}{the} MUX64 test setup.}
		\label{fig:massproductionsetup}
	\end{figure}
	The \textcolor{black}{schematic} of the MUX64 test setup is shown in figure \ref{fig:massproductionsetup}. The multiplexing functionality of \textcolor{black}{the} 64-to-1 is validated with \textcolor{black}{an} Agilent B2912A 2-channel precision source/measure unit \textcolor{black}{which} provides analogue voltage signals ($ \volt_{\mathrm{in}} $, A1 and $ 5\,\volt $, A2 in figure \ref{fig:massproductionsetup}). The MUX64 ON-channel is selected by 
	\textcolor{black}{six GPIO ports} on the isolated USB programmer board for lpGBT (UPL) \cite{Han_2022}. The input signals are connected to MUX64 via electromagnetic \textcolor{black}{relays}. A total of 64 relays are controlled by a computer interface in the test. The output signal is measured by \textcolor{black}{an} Agilent 34410A digital multimeter (V) on a $ 10\,\kilo\ohm $ load. 
	\begin{figure}[htb]
		\centering
		\begin{floatrow}
			\ffigbox[\FBwidth]{
				\centering
				\includegraphics[width=0.96\linewidth]{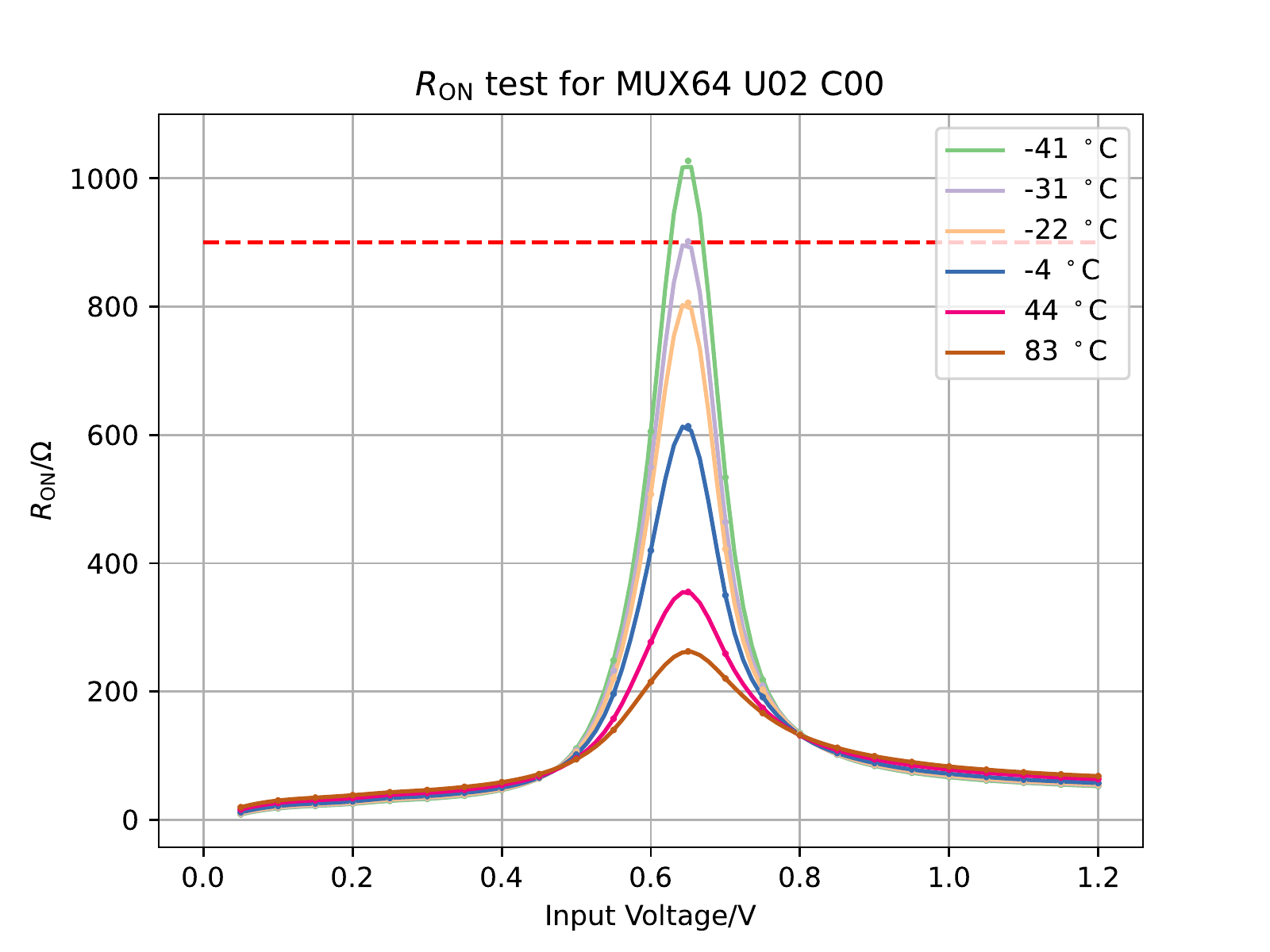}}{\caption{The $ R_{\mathrm{ON}} $ measured with a bare-die MUX64 (ch-0) is plotted versus input voltages in the temperature range of $ -41\,\degreecelsius $ to $ +85\,\degreecelsius $.}\label{fig:mux64temperatureron}}
			\quad
			\ffigbox[\FBwidth]{
				\centering
				\includegraphics[width=0.96\linewidth]{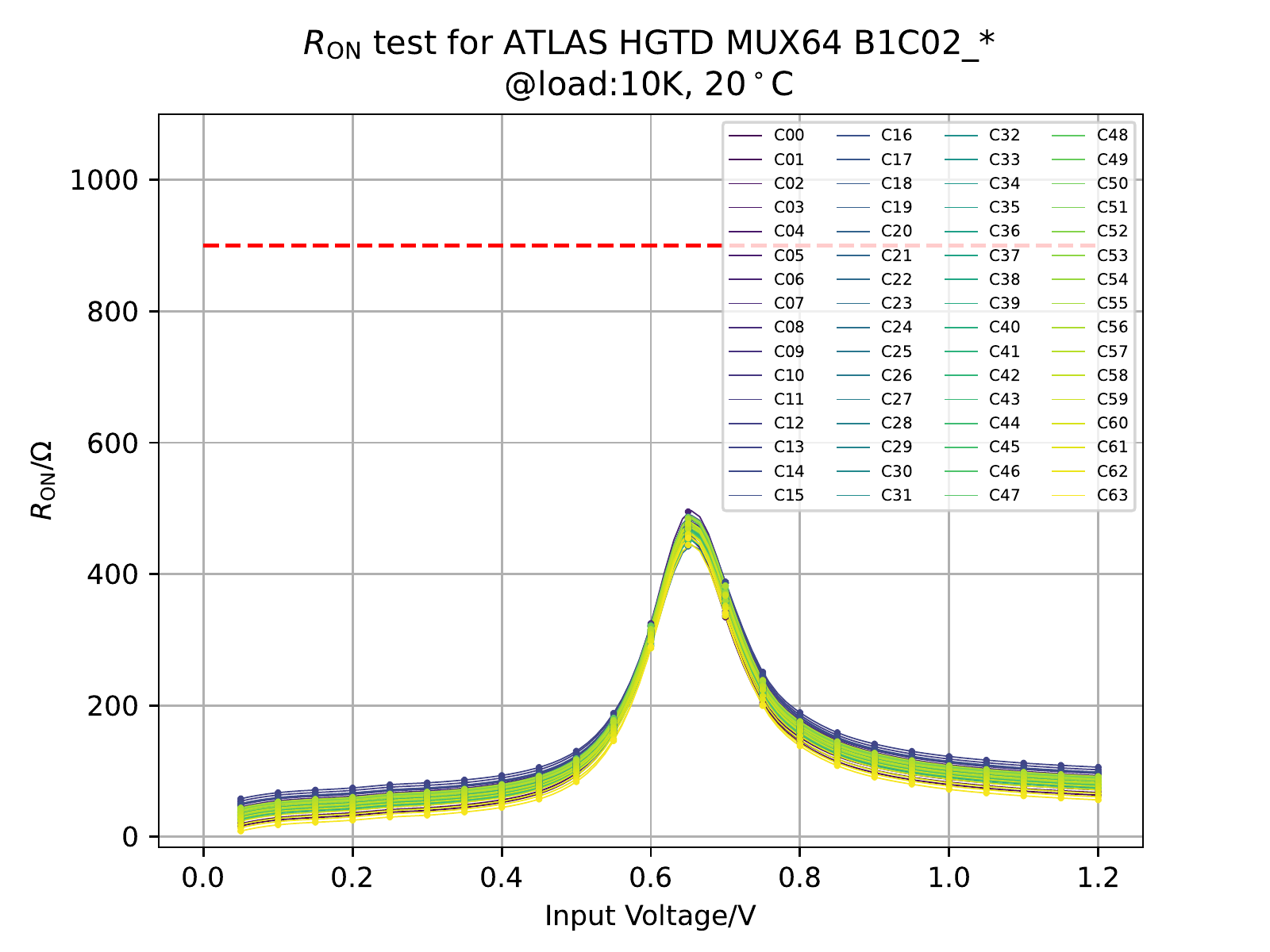}}{\caption{$ R_{\mathrm{ON}} $ vs input voltage curve of a \textcolor{black}{QFN88} packaged MUX64 at $ 20\,\degreecelsius $.}\label{fig:typicalonresistancecurve}}
		\end{floatrow}
	\end{figure}

	The $ R_{\mathrm{ON}} $ of a wire-bonded MUX64 channel is tested and plotted in figure \ref{fig:mux64temperatureron}. The input voltage varies from $ 0.05\,\volt $ to $ 1.20\,\volt $ in steps of $ 0.05\,\volt $ \textcolor{black}{at chosen temperatures ranged} from $ -41\,\degreecelsius $ to $ +85\,\degreecelsius $. The maximum $ R_{\mathrm{ON}} $ is measured at $ 0.65\,\volt $, which increases as the temperature decreases. At $ -31\,\degreecelsius $ the $ R_{\mathrm{ON}} $ peak is about $ 900\,\ohm $ which meets the design criteria. The rise of $ R_{\mathrm{ON}} $ at $ 0.65\,\volt $ is caused by the transmission gate switch \textcolor{black}{consisting} of a PMOS and \textcolor{black}{an} NMOS transistor, as is plotted in figure \ref{fig:mux64schameticunit}. The PMOS \textcolor{black}{affects} the $ R_{\mathrm{ON}} $ to increase nonlinearly with the input voltage while the NOMS \textcolor{black}{does} the opposite. The combined effects result \textcolor{black}{in} the distributions \textcolor{black}{are} shown in \textcolor{black}{figure} \ref{fig:mux64temperatureron} and \ref{fig:typicalonresistancecurve}.
	
	All the test samples were examined for the multiplexing function of every \textcolor{black}{channel} \textcolor{black}{at} room temperature. In figure \ref{fig:typicalonresistancecurve} the $ R_{\mathrm{ON}} $ of all 64 channels tested are plotted for a QFN88 packaged~ MUX64. \textcolor{black}{The OFF channel resistance is specified for $ 60\,\mega\Omega $. With all 64 channels turn off, the leakage to output is measured to be $ 25\,\nano\ampere $. The individual $ R_{\mathrm{off}} $ is sufficiently larger than the specified.}
	
	\begin{figure}[htb]
		\centering
		\subfloat[]{\includegraphics[width=0.45\linewidth]{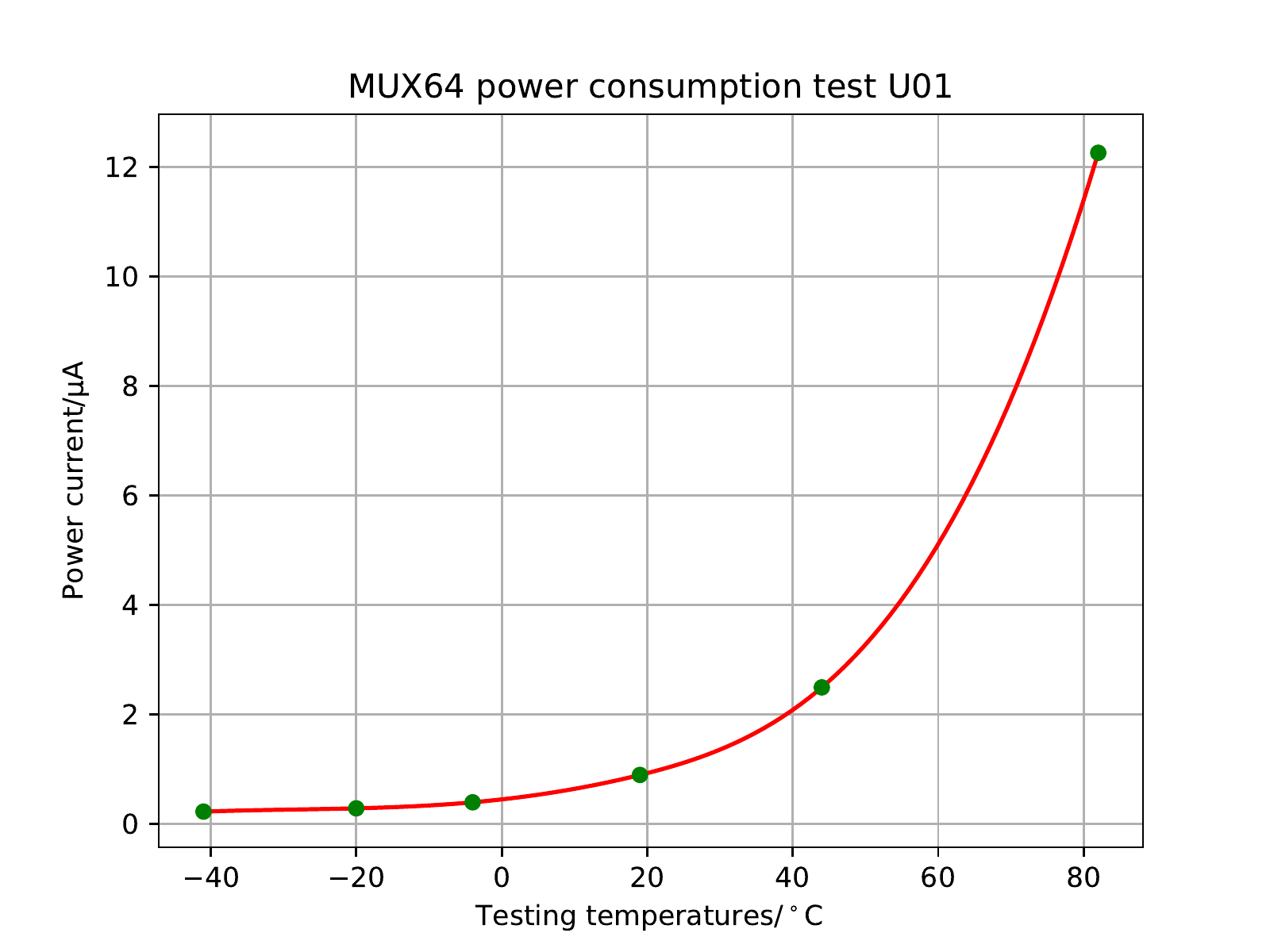}\label{fig:powerconsumption}}
		\quad
		\subfloat[]{\includegraphics[width=0.45\linewidth]{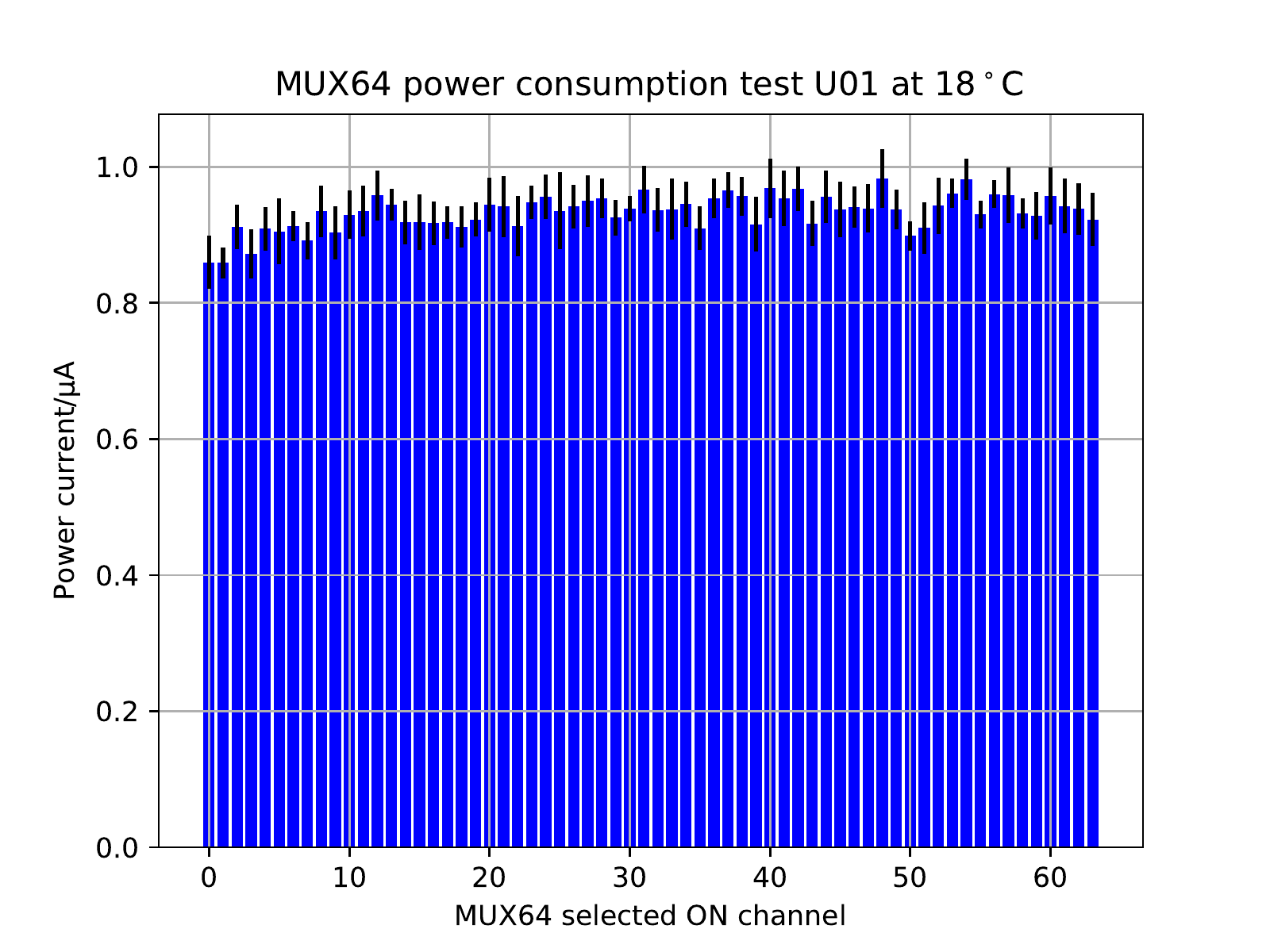}\label{fig:powerchannelbychannel}}\\
		\caption{The MUX64 power consumption was \textcolor{black}{measured} \textcolor{black}{for} (a) \textcolor{black}{the} power current versus operating temperature with a channel switched ON; (b) the current \textcolor{black}{at} $ 18\,\degreecelsius $ with one channel \textcolor{black}{being} switched ON in sequence. \textcolor{black}{The average at $ 18\,\degreecelsius $} is $ 0.93\pm0.06\,\micro\ampere $.}
		\label{fig:power_consumption_test}
	\end{figure}
	The power consumption of the MUX64 is examined for temperature dependence. In figure \ref{fig:powerconsumption}, the current of a MUX64 with a channel turned on is plotted versus temperature. The MUX64 power consumption decreases with operating temperature. At the detector operating temperature of $ -30\,\degreecelsius $, the power consumption at $ 1.2\,\volt $ is $ \sim0.1\,\micro\watt $. The uniformity of channels in operation is examined for the current variation, as is plotted in figure \ref{fig:powerchannelbychannel}. The current of MUX64 is measured at $ 18\,\degreecelsius $ with the channels switched ON in sequence. The average current measured is $ 0.93\pm0.06\,\micro\ampere $. The variance is less than 7\%.

	\begin{figure}[htb]
		\centering
		\subfloat[]{
			\includegraphics[width=0.35\linewidth]{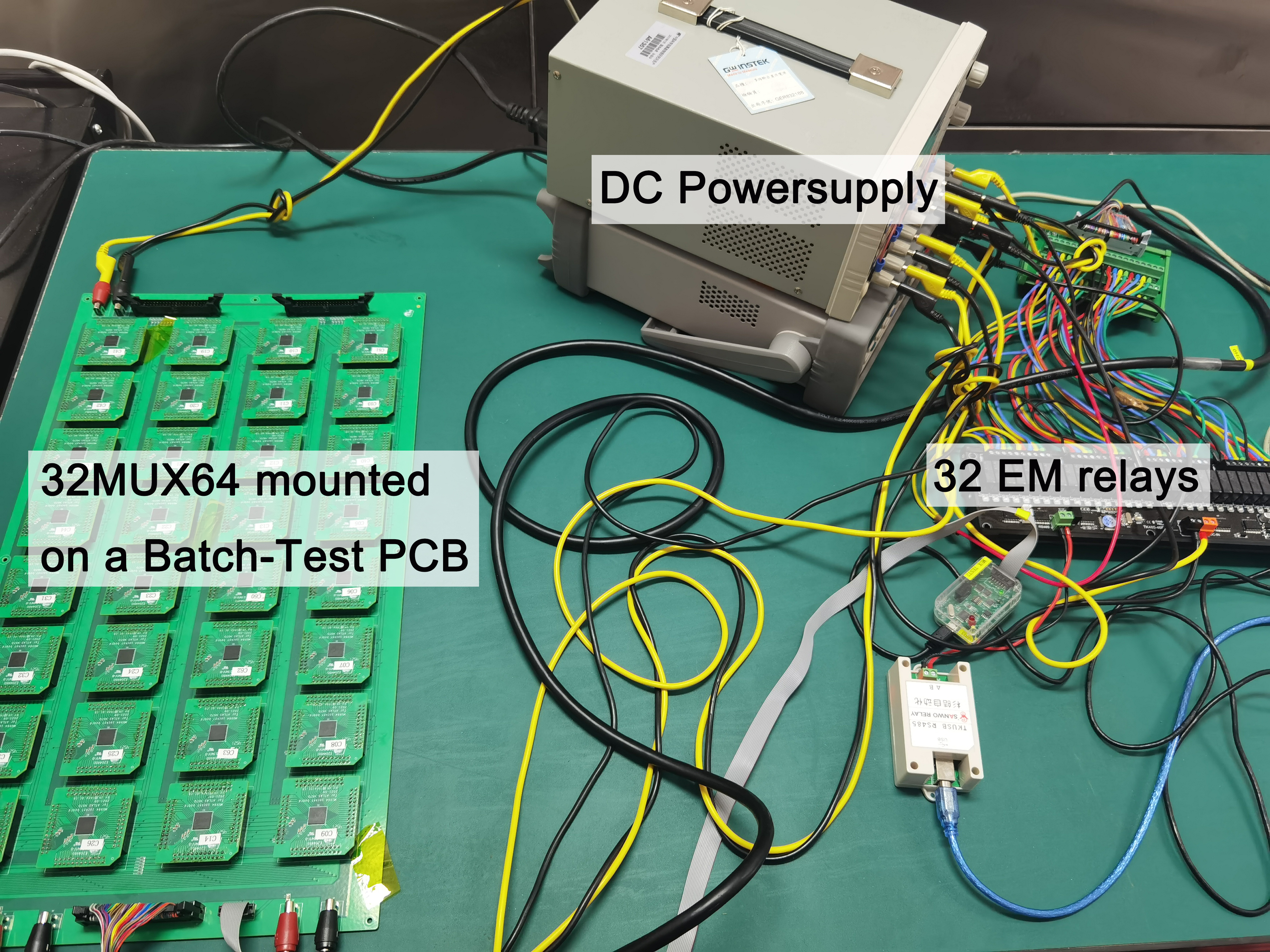}\label{fig:mux64reliabilitysetupa}}
		\quad
		\subfloat[]{
			\includegraphics[width=0.44\linewidth]{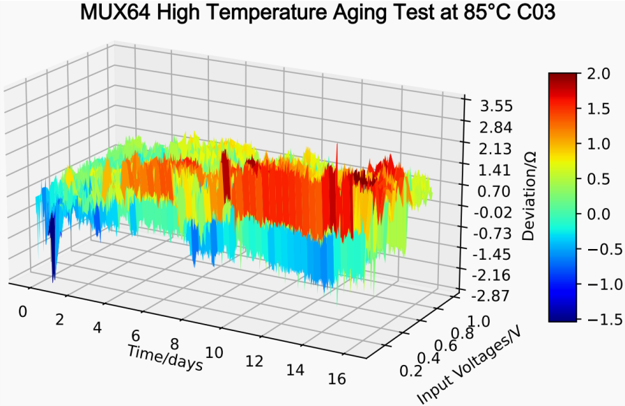}\label{fig:mux6416daysburninc03}} \\
		\caption{The burn-in test was conducted at $ 85\,\degreecelsius $ with (a) 32 QFN packaged MUX64s mounted on a Batch-Test PCB. Blue arrow: Batch-Test PCB; Red arrow: A QFN packaged MUX64. In (b), the deviation on $ R_{\mathrm{ON}} $ is plotted for a MUX64 being monitored during the 16 days burn-in period.}
	\end{figure}
	The reliability of MUX64 was further tested in burn-in. In the setup shown in figure \ref{fig:mux64reliabilitysetupa}, a total of 32 QFN88 packaged MUX64 chips were mounted on a Batch-Test (BT) board. The BT board has 64 signals \textcolor{black}{at} different \textcolor{black}{voltages each connected by fan-out cables} to all the MUX64s in \textcolor{black}{the} test. In the burn-in conducted at $ 85\,\degreecelsius $, all the test samples \textcolor{black}{withstood} well over 
	16 days. Figure \ref{fig:mux6416daysburninc03} shows the $ R_\mathrm{ON} $ deviation in 64 channels with different input voltages of a MUX64 monitored during the burn-in. The maximum deviations on $ R_\mathrm{ON} $ of all the MUX64s in all channels are smaller than $ 5\,\ohm $, which \textcolor{black}{demonstrates} the stability required for inputs to the ADC of the lpGBT. 
	
	
	\section{Conclusions and outlook}
	\label{sec:conclusions}
	We present the design and performance of MUX64. The prototype samples were tested and the result met design requirements. The burn-in test at 85 $ \degreecelsius $ with 32 chips shows negligible degradation over \textcolor{black}{16 days}. Quality assurance will be conducted on all the QFN packaged MUX64s with \textcolor{black}{thermal cycles} and channel-by-channel tests.
	
	The radiation tolerance of MUX64 is required. Irradiation test \textcolor{black}{was} carried out with the CSNS~ \cite{CN_first_PNS} $ 80\,\mega\electronvolt $ proton beam. The test samples had \textcolor{black}{withstood} a fluence of $ 2.10\times10^{15} \,(\mathrm{Si})\, n_{\mathrm{eq}}/\centi\meter^{2} $, which meets the requirement for HGTD. The tolerance with total ionizing doze (TID) is in progress using an X-ray irradiation facility.

	\acknowledgments
	We sincerely thank the Omega group (Omega/Ecole Polytechnique/CNRS, France) for their helpful assistance \textcolor{black}{in the} fabrication of the MUX64. This work is supported in part by \textcolor{black}{the} National Natural Science Foundation of China (NSFC) under \textcolor{black}{the} Contract $ \mathrm{No.}\,12122507 $. This work is supported by the National Natural Science Foundation of China ($ \mathrm{No.}\,11961141014 $)
%

	\bibliographystyle{JHEP}
	\bibliography{reference_mux64_paper}

\providecommand{\href}[2]{#2}\begingroup\raggedright\begin{thebibliography}{1}

\bibitem{CERN-LHCC-2020-007}
{\scshape ATLAS} collaboration, \emph{{Technical Design Report: A
  High-Granularity Timing Detector for the ATLAS Phase-II Upgrade}},
  CERN-LHCC-2020-007, ATLAS-TDR-031 (2020).

\bibitem{ATLAS_1502664}
{\scshape ATLAS} collaboration, \emph{{Letter of Intent for the Phase-II
  Upgrade of the ATLAS Experiment}},  CERN-LHCC-2012-022, LHCC-I-023 (2012).

\bibitem{PELLEGRINI201412}
G.~Pellegrini et~al., \emph{{Technology developments and first measurements of
  Low Gain Avalanche Detectors (LGAD) for high energy physics applications}},
  \href{https://doi.org/10.1016/j.nima.2014.06.008}{\emph{Nucl.~ Instrum. Meth.
  A} {\bfseries 765} (2014) 12}.

\bibitem{9507972}
C.~Agapopoulou et~al., \emph{{ALTIROC 1, a 25 ps time resolution ASIC for the
  ATLAS High Granularity Timing Detector}},  in \emph{2020 IEEE Nuclear Science
  Symposium and Medical Imaging Conference (NSS/MIC)}, pp.~1--4, 2020,
  \href{https://doi.org/10.1109/NSS/MIC42677.2020.9507972}{doi:
  10.1109/NSS/MIC42677.2020.9507972}.

\bibitem{Robles_Manzano_2022}
M.R.~Manzano et~al., \emph{{Design and testing results of a long flexible
  printed circuit for the {ATLAS} high granularity timing detector}},
  \href{https://doi.org/10.1088/1748-0221/17/06/c06001}{\emph{JINST} {\bfseries
  17} (2022) C06001}.

\bibitem{GBT_project}
P.~Moreira et~al., \emph{{The GBT Project}},  in \emph{{Proceedings of the
  Topical Workshop on Electronics for Particle Physics: Paris, France 21 - 25
  Sep 2009}}, pp.~342--346, 2009,
  \href{https://doi.org/10.5170/CERN-2009-006.342}{doi:
  10.5170/CERN-2009-006.342}.

\bibitem{Han_2022}
L.~Han et~al., \emph{{The isolated {USB} programmer board for {lpGBT}
  configuration in {ATLAS}-{HGTD} upgrade}},
  \href{https://doi.org/10.1088/1748-0221/17/03/c03030}{\emph{JINST} {\bfseries
  17} (2022) C03030}.

\bibitem{CN_first_PNS}
H.~Chen et~al., \emph{{China's first pulsed neutron source}},
  \href{https://doi.org/10.1038/nmat4655}{\emph{Nature Materials} {\bfseries
  15} (2016) 689}.

\end{thebibliography}\endgroup
	
\end{document}